\documentclass[twoside, a4paper, superscriptaddress, twocolumn, prl, aps, showpacs]{revtex4-1}

\usepackage[dvips]{graphicx}

\begin{document}

\title{Size limit for particle-stabilized emulsion droplets under gravity}

\author{J. W. Tavacoli}
\author{G. Katgert}
\affiliation{School of Physics and Astronomy, University of Edinburgh, Mayfield Road, Edinburgh EH9 3JZ, United Kingdom}
\author{E. G. Kim}
\affiliation{Institute for Basic Sciences, Changwon National University, Sarim-dong, Uichang-gu, Changwon-si, 641-773, Republic of Korea}
\author{M. E. Cates} \author{P. S. Clegg}
\affiliation{School of Physics and Astronomy, University of Edinburgh, Mayfield Road, Edinburgh EH9 3JZ, United Kingdom}

\date{\today}

\begin{abstract}
We demonstrate that emulsion droplets stabilized by interfacial particles become unstable beyond a size threshold set by gravity. This holds not only for colloids but supra-colloidal glass beads, using which we directly observe the ejection of particles near the droplet base. The number of particles acting together in these ejection events decreases with time until a stable acorn-like configuration is reached. Stability occurs when the weight of all remaining particles is less than the interfacial binding force of one particle. We also show the importance of the curvature of the droplet surface in promoting particle ejection.
\end{abstract}

\pacs{82.70.Kj; 82.70.Dd; 81.40.Np}

\maketitle

The use of particles as interfacial stabilizers for droplets and bubbles~\cite{Ramsden1903} has led to the discovery of extraordinary phenomena such as bubbles with long-term stability~\cite{Du2003}; `dry water'~\cite{Binks2006}; selectively permeable colloidosomes~\cite{Lee2008}; static bicontinuous gels~\cite{Tavacoli2011a}, and their templated networks~\cite{Sanz2009}.  All of these uses rely on the supramolecular size of the particles which prevents their thermal detachment from the interface. Such particles can then couple to external fields or field gradients in a way that molecular surfactants generally do not~\cite{Brown2012}. This idea has been used to create emulsions stabilized by superparamagnetic particles that can be dismantled using an applied magnetic field gradient~\cite{Melle2005} (see also~\cite{Hwang2010}). Particle-stabilized diving capsules also self-assemble to exhibit complex dynamics when a temperature gradient and a gravitational field are combined~\cite{Tavacoli2011b}.  

One external field, namely gravity, is present by default in almost all practical applications. The significance of this (or any other imposed field) is captured by a modified Bond number, $Bo$, which we define as the ratio of the field-induced body force $F_{G}$ on one particle, to the maximal capillary force, $F_{I}$, holding it on the interface: $Bo=a^{2}\Delta\rho g /\gamma (1-\cos{\theta})$. Here $a$ is the particle radius, $\Delta\rho$ is the density difference between the particles and the continuous phase, $g$ is gravity, $\gamma$ is the fluid-fluid interfacial tension and $\theta$ is the contact angle between the particle and the two liquids. A value of $Bo > 3/4$ indicates that gravity will pull an isolated particle off a flat interface. For many colloidal particles $a$ is on the micron scale while typically $\Delta\rho g / \gamma \approx 10^6$; hence one might imagine that gravity will never detach particles. However, this ignores the fact that body forces can be passed from one particle to another, causing collective instabilities to arise. Numerical explorations of these have identified two competing mechanisms: one in which a `keystone' particle is forced off the interface by the weight of particles resting on it, and another in which a rigidly jammed raft detaches as a block \cite{Kim2010,Kim2012}. 
\begin{figure}[htp]
\includegraphics[width=\columnwidth,]{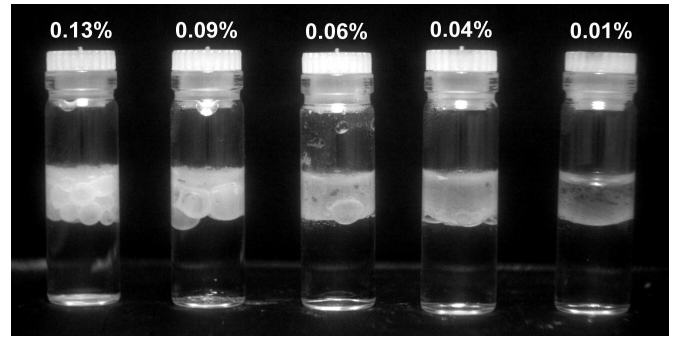}
\caption{A series of dodecane-in-water emulsions stabilized by melamine particles ($d = 3$\,$\mu$m) one day after formation. As the volume fraction of melamine is decreased (see labels) the droplets become larger and less stable. Vial diameter, 1.2\,cm.}\label{fig:melamine}
\end{figure}

In this Letter we explore experimentally the upper size limit for stable Pickering emulsion droplets in a gravitational field. We find that destabilization occurs via many particles acting to overcome the interfacial trapping of one or two particles i.e.\ the `keystone' mechanism. The behavior is also influenced by the curvature of the droplet. The number of particles found to be acting together is in reasonable agreement with expectations and can be used to determine the threshold droplet size.

We begin by observing particle-stabilized emulsions as the particle content is systematically reduced. We choose a system where buoyant oil droplets (dodecane) in water are stabilized by melamine particles (diameter 3\,$\mu$m, $\Delta\rho = 0.6$ g cm$^{-3}$) ; gravity pulls the particles down while the dispersed phase is bouyant. Emulsification was carried out using a vortex mixer and this results in polydisperse droplets. A series of samples is shown in Fig.~\ref{fig:melamine} one day after preparation. At the highest particle concentration a population of Pickering-stabilized droplets can be seen. For subsequent emulsions, with lower concentrations of particles, an overlayer of surplus dodecane is visible, together with a small number of stable droplets. The fact that one or more droplets remain stable in such samples is due to the spread of initial droplet sizes: most droplets were too large to remain intact. For the lowest two particle concentrations there are essentially no droplets present (although uncovered droplets can sometimes become trapped between the glass and the meniscus). Because during emulsification coalsecence proceeds until droplets have monolayer coverage, the role of decreasing particle concentration is to increase the size of the droplets at fixed coverage. Evidently we have chosen a system that forms stable Pickering emulsions and we have taken it beyond the size limit for stability. For the system addressed here, $Bo < 10^{-6}$: gravity is negligible for a single particle. Nonetheless our droplets become gravitationally unstable beyond a certain size.
\begin{figure}[htp]
\centering
\includegraphics[width=\columnwidth,]{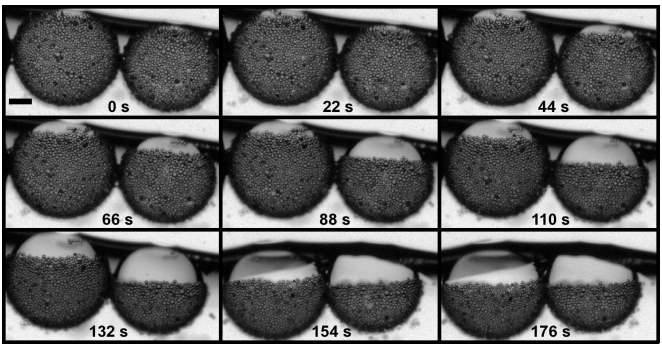}
\caption{A time sequence of frames, beginning top left, of a pair of Pickering droplets as beads are pulled from the interface by gravity. Timing of each frame is shown starting from when the droplets make contact with the meniscus. Shedding slows until the (smaller) droplet begins to resemble a stable \textit{acorn} in the last few frames (bottom row). Scale bar 200\,$\mu$m.}\label{fig:acorn}
\end{figure}

To resolve what is happening when particle-stabilized droplets become unstable we use larger particles whose individual motions can be tracked. A monolayer of polydisperse glass beads (average radius = 10 $\mu$m, $\Delta\rho = 1.6$ g cm$^{-3}$, with a grafted polyethylene glycol layer to suppress interparticle attraction) was trapped on the interface of an octane droplet (size $\approx$ 0.8\,mm; approx 5000 particles per droplet) in a continuous phase of distilled water. Such Pickering droplets were prepared by hand-mixing the three components and were found to be stable for months at least. From observations of isolated particles we find $\theta = 10 - 15^{\circ}$ and hence $Bo \approx 10^{-3}$. To destabilize the droplet we reduced the interfacial tension by transferring the droplets to a continuous phase containing 7\,mM sodium dodecyl sulfate surfactant (SDS) (see~\cite{Supplementary}; SDS was chosen as it has been shown not to adsorb onto similar particles). In the presence of SDS we find that the interfacial tension decreases while the contact angle remains unchanged giving $Bo \approx 10^{-2}$; thus the ejection of isolated beads is still not anticipated. The aqueous surfactant phase was positioned underneath an overlying layer of octane and both liquids were contained in a glass cell that was silanized to control the shape of the meniscus. We observed our Pickering droplets via a CCD camera (frame rate 5\,Hz). 

These droplets, stable at $Bo \approx 10^{-3}$, began to expel beads from their interfaces at $Bo\approx 10^{-2}$, Fig.~\ref{fig:acorn}. As individual beads left the interface there was reorganization of those that remained trapped. The rate of ejection of beads slowed with time until eventually a stable configuration was reached (Fig.~\ref{fig:acorn} bottom row, smaller droplet). What began as a fully covered Pickering droplet finally resembled an acorn: the octane droplet was coated with a shell of particles only around the lower hemisphere. (The fact that some beads always remain undetached confirms their affinity for the interface even in the presence of SDS.) The droplets did eventually coalesce with the overlying octane layer; however, in the majority of cases this did not occur until beads had ceased to fall off \cite{foot1}.  Importantly, we observed similar particle ejection behavior for other Pickering droplets, including cases without surfactant (see~\cite{Supplementary}). 

Our experimental setup allows us to quantify the evolving configurations of particles on the droplet surface and hence the collective behavior leading to their ejection. This was achieved by extracting particle coordinates (frame rate 9\,Hz) using the image analysis software ImageJ and IDL~\cite{ImageJ}. From the resulting tracks we calculated a velocity correlation function, $g(r) = \langle \mathbf{v}(r') . \mathbf{v}(r' + r) \rangle / \langle v^2 \rangle,$ where $\mathbf{v}(r')$ is the velocity in the image plane of a central reference particle and $\mathbf{v}(r' + r)$ is the velocity in the image plane of a particle separated from the first by a direct distance $r$. Note that $g(r)$ was determined for each recorded frame (save the first) and the average $\langle \dots \rangle$ was performed over all the other particles in the frame. In any given ejection event, a reference bead was chosen in the center of a mobile patch. This facilitated the calculation of a patch radius, $R_g$, from the decay of $g(r)$. We use this to estimate the patch area, $\pi R_g^2$, and hence the bead population in a patch. 

\begin{figure}[htp]
\centering
\includegraphics[width=\columnwidth,]{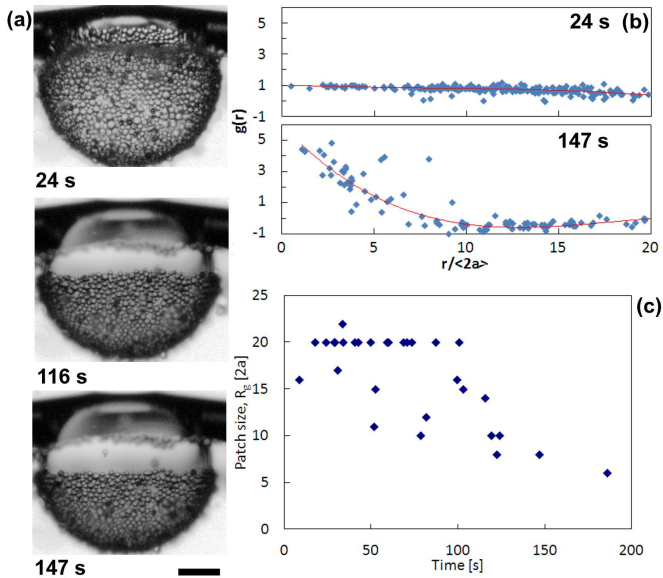}
\caption{(a) Images and (b) velocity-velocity correlation functions for different times as the droplet is destabilized by gravity. There is a clear decrease in length over which velocity correlations are maintained with time. Scale bar 200\,$\mu$m. (c) The radius of patches of particles, $R_g$, deduced from the velocity-velocity correlation functions. Initially sheet-like motion is responsible for ejection events (i.e. the patch is the size of the droplet), after about 100\,s the patch size decreases continuously until bead ejection finally ceases.}\label{fig:PatchSize}
\end{figure}

The average size of these patches reduces with time as demonstrated in Fig.~\ref{fig:PatchSize}. The changing character of the ejection dynamics is best explained in terms of two limiting pathways, one dominant at early stages (Fig.~\ref{fig:PatchSize}a, 24\,s) and the other later (Fig.~\ref{fig:PatchSize}a, 147\,s). At early stages, the velocities of all the beads are highly correlated, and a coherent downward motion of all particles in unison is seen. Hence $g(r)$ is a positive constant. By contrast, late stage loss results in the localized rearrangements of particles in patches and here, $g(r)$ is a decaying function. This second mechanism replaces the first after about 100~sec. These results obtained from the direct analysis of  velocity correlations are supported by an alternative method where the patch sizes were deduced from a four-point correlation function, $\chi^4$ (see~\cite{Supplementary}); this is a well known procedure to obtain dynamical length scales from experiments on granular media.

Our analysis of the bead motions strongly suggests that particle ejection is a collective effect in which a certain minimum number $N_m$ of beads participate. For glass beads on these octane droplets in SDS/water we find that the limiting patch width is $\simeq 10$ beads so that $N_m \simeq 100$. No further ejection is seen once the total number of particles on a droplet falls below $N_m$. Notably we do not see detachment of large groups of particles simultaneously; instead each ejection event leads to the detachment of only one or two particles.

Computer simulations of large Pickering droplets under gravity suggest two distinct mechanisms for particle ejection. In~\cite{Kim2012}, hard sphere particles with neutral wetting ($\theta = 90^{\circ}$) were modelled, with initial areal density $\alpha \simeq 0.53$). Results were broadly consistent with a maximum droplet radius found by balancing the body force on a lower hemisphere of particles against the interfacial force across the equator of a droplet. Such a droplet then has $N_c\sim Bo^{-2}$ particles. In this failure mode (see also~\cite{Dominguez2008}) a jammed raft of particle detaches en bloc rapidly leading to droplet fission. However it was also reported in~\cite{Kim2012} that on reducing the interfacial absorption radius of the particles by a factor two while maintaining their hard core repulsion radius, a different failure mode was seen. In this mode, a single `keystone' particle gets ejected under the cumulative body force of a patch of particles bearing down upon it. The simulated geometry closely mimics the effect of reducing the contact angle $\theta$ \cite{foot2}: both reduce the force needed to detach a single keystone particle but leave unchanged that for the collective detachment of a jammed raft. The droplet stability limit set by this failure mode is governed by $N_c\simeq N_m\sim Bo^{-1}$; here $N_m$, as previously identified, is the size of a patch whose total body force just overcomes the interfacial trapping of a single bead.

Our experimental studies on glass beads address systems with relatively small $\theta$. The observed behavior decisively points to the keystone-like instability as the limiting failure mode in the range of parameters studied, first because of the observation that $N_m \simeq Bo^{-1} \simeq 100$ (rather than $N_m\simeq Bo^{-2} \simeq 10^4$ for the raft mode), and second because we directly observe that ejection proceeds by successive detachment of isolated particles rather than of large rafts. 
 
In most situations, practitioners are interested in sizes of droplets and not numbers of particles. The above reasoning can be recast to give the limiting droplet size for Pickering emulsions under gravity (or, mutatis mutandis, any other body force) as 
\begin{equation} \label{criticalradius}
D_{max} \simeq \sqrt{\gamma (1 - \cos{\theta}) / \Delta\rho g}.
\end{equation}
which is further justified in~\cite{Supplementary}. This limiting droplet size takes the form of a capillary length. For given interfacial properties and densities, it sets a practical formulation ceiling that cannot be overcome by varying the particle size $a$, which does not enter. As we have shown (Fig.~\ref{fig:melamine}), when increasingly large Pickering emulsion droplets are fabricated using high density colloidal particles, a critical size is reached whereupon stability against coalescence is lost. For dodecane droplets, stabilized by melamine particles, in water, we find $D_{max} \approx 3$\,mm in good agreement with our observations. Furthermore Eqn.~1 is consistent with the size of the stable POD described in ref.~\cite{Tavacoli2011b}. The limiting size $D_{max}$ can, however, be overcome by strong interparticle attractions (which present different challenges for practical emulsification). Our findings may also explain results by Melle et al. \cite{Melle2005} who, via a magnetic field gradient, were able to remove paramagnetic particles from the base of buoyant Pickering droplets, despite generating forces 10$^{6}$ times lower than $F_I$ on individual particles. Intriguingly, there were $\approx$10$^{6}$ particles on their unstable droplets.

\begin{figure}[htp]
\centering
\includegraphics[width=\columnwidth,]{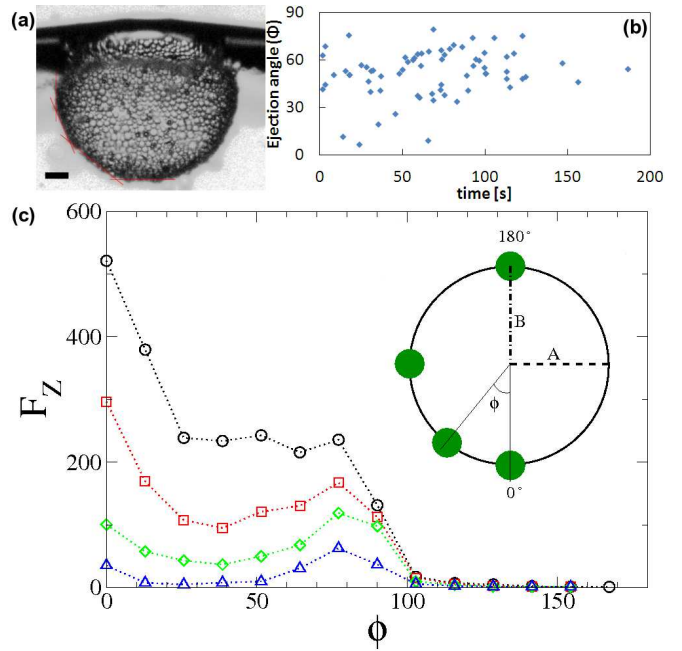}
\caption{(a) Buoyant droplets covered with glass beads are non-spherical. The base becomes flattened with somewhat vertical sides. Scale bar 100\,$\mu$m; (b) The experimental elevation angle for ejection events as a function of time. Initially some of the events occur around the base of the droplet ($\phi \approx 0^{\circ}$); subsequently ejection occurs only at higher elevations ($\phi > 0^{\circ}$); (c) Calculated vertical force on simulated interfacial particles due to gravity as a function of the elevation on the droplet surface. Values for spherical and distorted droplets with various values of flattening ($f$) are shown: black (0), red (0.19), green (0.36), and blue (0.5). As $f$ increases the maximum force, $F_z$, moves from $\phi = 0^{\circ}$ to $\phi = 78^{\circ}$.}\label{fig:distorted}
\end{figure}

To underline the collective nature of the bead ejection events we finally explore how they can be influenced by droplet shape. Our large, buoyant, droplets are distorted away from spherical as they are compressed against the meniscus \cite{foot1}. For example, the droplet shown in Figure~\ref{fig:distorted}a has a straightened central region, where beads are essentially stacked above each other, and a compressed base. There is a high curvature area where these two regions meet and it is here where we observe bead ejection to be mainly localized. In Fig.~\ref{fig:distorted}b we plot the elevation angle, $\phi$, of ejection events versus time beginning from when a droplet makes contact with the meniscus. Initially, particles are ejected at a range of elevations but later, as the distortion settles in, ejection events cease at the lowest point, $\phi = 0^{\circ}$. In the absence of particle friction, a vertical body force cannot be transmitted between particles within a purely horizontal layer so that presumably the largest cumulative detachment forces arise in the curved region connecting the flattened base with the vertical sides. That the position of bead ejection is dependent on the droplet shape is emphasized when we notice that in the short period when our droplets are climbing to the water-oil meniscus, during which the droplets are elongated vertically, particles are seen to be shed at the very base of the droplet (compare \cite{Kim2012}). The slightly slower approach to a quiescent state for the bigger droplet in Fig.~\ref{fig:acorn} might be caused by the ring where beads are lost growing more slowly ($\sim D$) with droplet size than the number of beads to be shed ($\sim D^2$) leading to a `bottleneck' at large droplet sizes.

To corroborate these observations we have calculated the vertical component of the force for a simulated population of particles on a droplet surface, Fig.~\ref{fig:distorted}c (the particles are square-packed: size is varied to maintain contacts). The droplet flattening, $f=(A-B)/A$, (see inset) is systematically increased. For a spherical droplet the maximum force is right at the base ($\phi = 0^{\circ}$) but as the droplet becomes increasingly oblate the maximum force moves away from the base and for sufficiently large distortions occurs at $\phi = 78^{\circ}$.

In conclusion, we find that buoyant Pickering emulsion droplets which employ particles that are more dense than the continuous phase become gravitationally unstable beyond a certain droplet size. We have demonstrated, using large beads trapped at droplet interfaces, that bead ejection requires correlated motion by a \textit{patch} of particles. We find that the smallest patch responsible for a bead ejection event contains roughly $Bo^{-1}$ particles; this is the number of particles whose weight must be combined to overcome the interfacial trapping of a single `keystone' particle. This can be restated as a maximum stable droplet size (Eqn.~\ref{criticalradius}) which depends on material properties and field strength but not the size of the stabilizing particles. The observed mechanism for droplet failure may however depend on the contact angle $\theta$ in which case a different result might apply as neutral wetting ($\theta = 90^\circ$) is approached \cite{Kim2012}. Future experiments to explore this dependence would be very worthwhile.

The proposed instability mechanism applies not only to gravity but to any external field resulting in a body force acting equally on all interfacial particles. This work is thus of broader relevance to understanding the stability of Pickering emulsions and its manipulation using external fields. E.g.\ it may help explain why emulsions stabilized by superparamagnetic particles could be destabilized using relatively small magnetic field gradients~\cite{Melle2005}.

Sadly, GK passed away before this manuscript was completed. Work funded by EPSRC EP/E030173/01; PSC and MEC are funded by the Royal Society.


\begin{thebibliography}{100}
\bibitem{foot1} That the droplets remain stable for a significant time at the meniscus, even when there were no particles present to prevent octane-octane contact, is due to surfactant at the octane-water interface.
\end{thebibliography}
\end{document}